\documentclass[10pt,letterpaper]{article}

\usepackage{gensymb}

\usepackage[top=0.85in,left=2.25in,footskip=0.75in,marginparwidth=2in]{geometry}

\usepackage[utf8]{inputenc}

\usepackage{cite}

\usepackage{nameref,hyperref}

\usepackage[right]{lineno}

\usepackage{microtype}
\DisableLigatures[f]{encoding = *, family = * }

\raggedright
\usepackage{changepage}

\usepackage[aboveskip=1pt,labelfont=bf,labelsep=period,singlelinecheck=off]{caption}

\makeatletter
\renewcommand{\@biblabel}[1]{\quad#1.}
\makeatother

\usepackage{lastpage,fancyhdr,graphicx}
\usepackage{epstopdf}
\fancyheadoffset[L]{2.25in}
\fancyfootoffset[L]{2.25in}

\usepackage{color}

\definecolor{Gray}{gray}{.25}

\usepackage{graphicx}

\usepackage{sidecap}

\usepackage{wrapfig}
\usepackage[pscoord]{eso-pic}
\usepackage[fulladjust]{marginnote}
\reversemarginpar

\begin{document}
\vspace*{0.35in}

% title goes here:
\begin{flushleft}
{\Large
\textbf\newline{Untethered microbot powered by giant magnetoelastic strain}
}
\newline
% authors go here:
\\
Yukun Xia\textsuperscript{1},
Jikun Wang\textsuperscript{2},
Ruisen Yang\textsuperscript{2},
Tongqing Lu\textsuperscript{2*},
Tiejun Wang\textsuperscript{2},
Zhigang Suo\textsuperscript{1*}
\\
\bigskip
\bf{1} John A. Paulson School of Engineering and Applied Sciences, Harvard University, Cambridge, MA 02138, United States
\\
\bf{2} State Key Lab for Strength and Vibration of Mechanical Structures, International Center for Applied Mechanics, Department of Engineering Mechanics, Xi’an Jiaotong University, Xi’an 710049, China
\\
\bigskip
* tongqinglu@mail.xjtu.edu.cn, suo@seas.harvard.edu

\end{flushleft}

\section*{Abstract}

\quad \quad Magnetic elastomers deform under a magnetic field, working as a soft actuator. Untethered microbot made of magnetic elastomers have great potentials in performing medical tasks inside human body as mini doctors. However, the lack of a highly deformable and efficient actuator strongly limits the development of magnetic microbots. In this paper, we developed an actuator of magnetic elastomer capable of large deformation strain by harnessing magnetic pull-in instability. We design three prototypes of untethered microbots by using this actuator: a robot grips objects with a large rotating angle, a robot navigates in a fluid channel and delivers drug, and a robot self-accelerates and quickly jumps up. The pull-in mechanism enables a new type of soft robot with fast response, remote wireless control and extremely high power density. 

% now start line numbers
%\linenumbers

% the * after section prevents numbering
\section*{Introduction}

\quad \quad Subject to an external magnetic field, magnetic particles embedded in a soft matrix tend to rotate along the field direction and meanwhile interact with one another. This mechanism has been used as soft magnetic actuators for about two decades. Magnetic elastomers can be actuated either by a permanent magnet [1-4] or a controllable gradient magnetic field from an electromagnet [5-7]. Driven by a gradient magnetic field, the magnetic elastomer tends to move towards the magnetic field source. Driven by a uniform magnetic field, lined-up magnetic particles can bend locally, leading to various deformation modes. For example, Kim et al. patterned the magnetic particle chain in a photocurable resin and achieved snake-like movement under a uniform magnetic field [8]. Huang et al. fabricated self-assembly membranes, which could reconfigure its shape in a magnetic field as a microbe-like swimmer [9]. Schmauch et al. also designed the magnetic membranes to function as a lifter, valve and accordion [10]. Another recent [11] printed magnetic ink into heterogeneous membrane, which could transform under magnetic field into complicated 3D structures through exquisite design and control. However, the actuation strain in all the above actuators is very small. No soft actuators capable of large magnetic actuation strain under a uniform magnetic field have been demonstrated so far.

\quad \quad The high magnetic field is harmless to human body, leading to the great development of the technique of nuclear magnetic resonance in the field of medical diagnosis. In recent years, it becomes a hotspot topic to carry out precise surgical tasks inside human body with a magnetic microbot [12-15]. Magnetic microbots can be injected into human body and move by an external magnetic field to the target area to perform tasks, such as navigating in blood vessels to deliver drug [16-18], cleaning out block [19], executing local hyperthermia [20] and marking pathologies [21]. For example, Jeong et al. and Ullrich et al. respectively conducted vivo experiments to manipulate a magnetic unit moving in the blood vessel of a pig and the eye of a rabbit [22, 23]. Felfoul et al. used the magnetic bacteria as the substrate of the hybrid microbot, and controlled the movement of these bacteria with magnetic field [24]. However, these microbots are still far from being an effective medical tool, due to the lack of a deformable and efficient magnetic actuator.

\quad \quad In this work, we revealed the mechanism of magnetic pull-in instability of magnetic elastomers. We fabricated a soft magnetic actuator capable of reversible 60\% actuation strain in a uniform magnetic field by harnessing the localized pull-in instability. As demonstrations, we designed three robots with simple geometries to show the powerful functions of soft magnetic actuators. The first robot can grip object with a large rotating angle, the second robot can navigate in a fluid environment and deliver drug at the designated location, and the third robot can quickly jump up under a magnetic field, raising its gravity center several times higher.

\section*{Large magnetoelastic strain}

\subsection*{Pull-in instability of two magnetic particles} 

\quad \quad Consider two spherical iron particles embedded in an elastomeric matrix. The two particles are placed along the vertical direction. A uniform magnetic field $H$ is applied in the same direction (Fig. 1a). The two particles are magnetized and the distance

\marginpar{
\vspace{.7cm} % adjust vertical position relative to text with \vspace{} - note that you can enter negative numbers to move margin caption up
\color{Gray} % this gives caption a grey color to set it apart from text body
\textbf{Figure 1. Magnetic pull-in instability in a uniform magnetic field.} % note that \ref{fig1} refers to the corresponding wrapfigure
(a) Two iron particles are embedded in a bulk of elastomer. When the applied magnetic field is small, the magnetic force ($F_{magnetic}$) is balanced with the elastic force ($F_{elastic}$). When the magnetic field reaches a critical value, the two particles pull in together, squeezing the elastomer in between dramatically. (b) The normalized magnetic field as a function of the distance change of the two particles. 
}
\begin{wrapfigure}[24]{l}{75mm}
% the number in [] of wrapfigure is optional and gives the number of text lines that should be wrapped around the text. Adjust according to your figures height
\includegraphics[width=75mm]{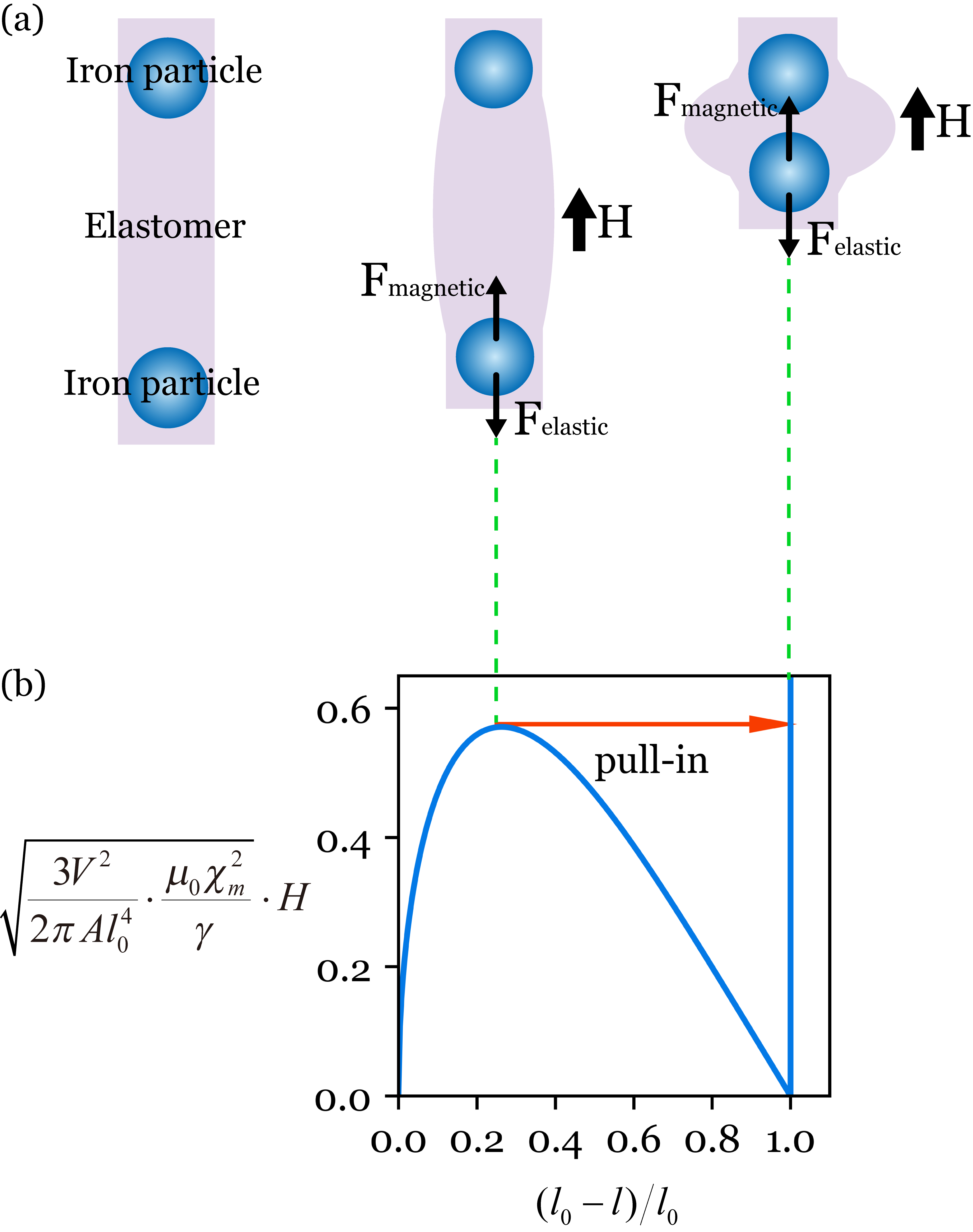}
\captionsetup{labelformat=empty} % makes sure dummy caption is blank
\caption{} % add dummy caption - otherwise \label won't work and figure numbering will not count up
\label{fig1} % use \ref{fig1} to reference to this figure
\end{wrapfigure} % avoid blank space here

between them change from $l_0$ in the absence of magnetic field to $l$. The equilibrium is reached by balancing the elastic force $F_e$ and the magnetic force $F_m$. Gravity is neglected. We assume the distance between the two particles $l$ is large compared to the radius of particle $R$ and take the two particles as two dipoles. To a first approximation, the magnetic force is roughly $F_m \approx 3\mu_0 m^2 / (2\pi l^4)$, where $\mu_0$ is the permeability of vacuum, and $m$ is the magnetic dipole moment of each particle. We further simplify that the relation between the dipole moment and the applied field is linear $m=V\chi_m H$, where $V$ is the volume of particle and $\chi_m$ is the effective magnetic susceptibility. The elastomeric matrix is taken to be a neo-Hookean material so that the elastic force is estimated as $F_e = A \gamma [l/l_0 - (l/l_0)^{-2}]$, where $A$ is the area of section in the undeformed state and $\gamma$ is the shear modulus of the matrix. Force balance gives that

\begin{equation}
\sqrt{\frac{3V^2}{2\pi Al_0^4} \cdot \frac{\mu_0 \chi_m^2}{\gamma2}} \cdot H = \sqrt{(l/l_0)^5 - (l/l_0)^3}
\end{equation}

\quad \quad The normalized applied magnetic field as a function of compressive strain $(l_0-l)/l_0$ is plotted in Fig. 1b. When the magnetic field is small, the matrix is stiff enough to resist the magnetic force and the elastic deformation is small. As the magnetic field increases, the distance between the two particles decreases and the magnetic force increases much more rapidly than the elastic force. Once the distance reaches a critical value, the two particles suddenly pull-in and squeeze the matrix in between significantly. The critical condition corresponds to the peak point in Fig. 1b. Once the geometric parameters $A$, $l_0$, $V$ are given, the critical condition to trigger the pull-in instability is determined by the competition between elastic modulus $\gamma$ and the magnetic Maxwell stress $\frac{1}{2}\mu_0 \chi_m^2 H^2$. When the matrix is soft and the magnetic susceptibility of particles is high, the magnetic pull-in instability is prone to happen. The maximum compressive displacement is roughly the initial distance between the two particles, as indicated by the vertical line in Fig. 1b.

\subsection*{Experimental demonstration of magnetic pull-in}

\quad \quad We synthesized a transparent polyacrylamide hydrogel as the soft matrix. Five cylindrical particles made of pure iron were embedded in the matrix (Fig.2a). The

\marginpar{
\vspace{.7cm} % adjust vertical position relative to text with \vspace{} - note that you can enter negative numbers to move margin caption up
\color{Gray} % this gives caption a grey color to set it apart from text body
\textbf{Figure 2. A linear magnetoelastic actuator.} % note that \ref{fig1} refers to the corresponding wrapfigure
(a) Five aligned iron particles embedded are embedded in a soft hydrogel matrix in the absence of magnetic field; (b) Subject to a uniform vertical magnetic field the particles pull in together, generating a linear actuation strain about 60\%.
}
\begin{wrapfigure}[16]{l}{75mm}
% the number in [] of wrapfigure is optional and gives the number of text lines that should be wrapped around the text. Adjust according to your figures height
\includegraphics[width=75mm]{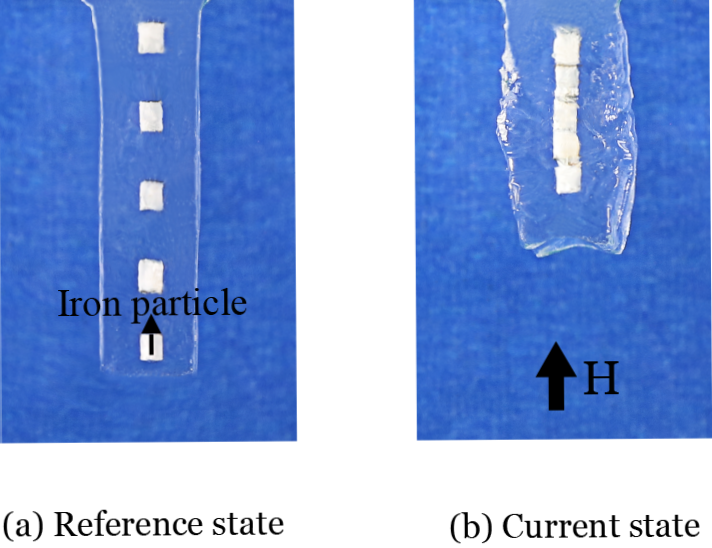}
\captionsetup{labelformat=empty} % makes sure dummy caption is blank
\caption{} % add dummy caption - otherwise \label won't work and figure numbering will not count up
\label{fig1} % use \ref{fig1} to reference to this figure
\end{wrapfigure} % avoid blank space here

particles were painted in white color. The size of hydrogel was $12mm \times 9mm \times 55mm$, and the size of each particle was $3mm \times 3mm \times 4mm$. When a small uniform magnetic field was applied, the relative displacement between particles was unapparent. When the magnetic field was increased to about 0.4T, the particles suddenly pulled in, producing a large deformation (Fig.2b). The length between the top and the bottom particle changed to roughly 40\% of its initial value. When the magnetic field was removed, the actuator returned to the initial state. The shear modulus of the hydrogel matrix is about 2kPa. The magnetic susceptibility of particles $\chi_m$ is roughly 3, accounting for the demagnetization effect [25]. The critical magnetic field for pull-in is estimated to be 0.3T, according to Eq.1. The comparison between the theoretical prediction and the experiments is reasonable. The discrepancy is possibly due to the simplification of geometry and the linear magnetization relation. 

\subsection*{Placement of magnetic particles}

\quad \quad Consider two magnetic particles placed in a soft matrix along a direction which is not parallel with the external magnetic field. Set the origin of the coordinate on one particle, and the other particle has the relative coordinate $(x,y)$. When the distance between the particles is large compared to the particle size, the potential energy of the two particles can be estimated as $\displaystyle -\frac{\mu_0 m^2}{4\pi}\frac{2y^2 - x^2}{(x^2 + y^2)^{\frac{5}{2}}}$ [26, 27]. When the external magnetic field is along the vertical direction, the two particles are subject to attractive force if they are placed vertically, are subject to repulsive force if placed horizontally, and are subject to zero force if they are placed with an inclined angle (Fig. 3a). \\

\marginpar{
\vspace{.7cm} % adjust vertical position relative to text with \vspace{} - note that you can enter negative numbers to move margin caption up
\color{Gray} % this gives caption a grey color to set it apart from text body
\textbf{Figure 3. Placement of magnetic particles greatly affects the overall response.} % note that \ref{fig1} refers to the corresponding wrapfigure
(a) Under a uniform vertical magnetic field, particle 1 and particle 2 are subject to attractive force if they are placed vertically, subject to repulsive force if they are placed horizontally. (b) and (c) two commonly used magnetic elastomers: random distribution and chain-like distribution.
}
\begin{wrapfigure}[10]{l}{100mm}
% the number in [] of wrapfigure is optional and gives the number of text lines that should be wrapped around the text. Adjust according to your figures height
\includegraphics[width=100mm]{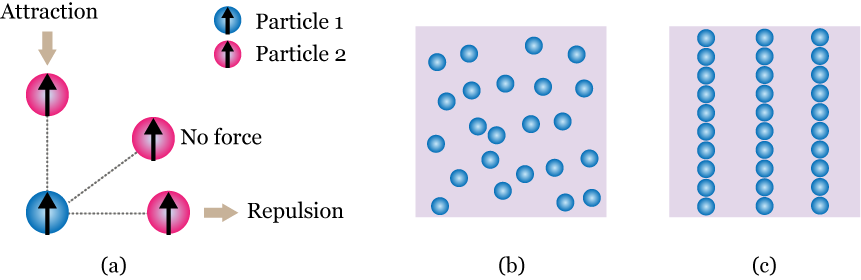}
\captionsetup{labelformat=empty} % makes sure dummy caption is blank
\caption{} % add dummy caption - otherwise \label won't work and figure numbering will not count up
\label{fig1} % use \ref{fig1} to reference to this figure
\end{wrapfigure} % avoid blank space here(a) Under a uniform vertical magnetic field, particle 1 and particle 2 are subject to attractive force if they are placed vertically, subject to repulsive force if they are placed horizontally. (b) and (c) two commonly used magnetic elastomers: random distribution and chain-like distribution. \\

Consequently, if magnetic particles are randomly placed in the soft matrix (Fig. 3b), the attractive force and repulsive force offset each other and the achievable actuation strain is very small [28]. Magnetorheological elastomer, cured in a magnetic field to form the chain-like particle distribution (Fig. 3c), can change stiffness under a magnetic field [29, 30]. The structure avoids the force offset but is not suitable for strain actuation, because the distance between adjacent particles is too small and no space is left for generating compressive strain.

\quad \quad We design a cubic lattice distribution of magnetic particles in a soft matrix which can avoid the force offset and undergo large deformation. We assembled 27 iron particles into a silicone matrix (ecoflex-0030) (Fig.4a). The particles were placed as 

\marginpar{
\vspace{.7cm} % adjust vertical position relative to text with \vspace{} - note that you can enter negative numbers to move margin caption up
\color{Gray} % this gives caption a grey color to set it apart from text body
\textbf{Figure 4. Spatially equidistant distribution.} % note that \ref{fig1} refers to the corresponding wrapfigure
(a) In reference state, 27 Iron particles were uniformly placed in a soft matrix to form a cubic lattice. (b) In current state, under a vertical magnetic field, the particles pull in together. 
}
\begin{wrapfigure}[14]{l}{75mm}
% the number in [] of wrapfigure is optional and gives the number of text lines that should be wrapped around the text. Adjust according to your figures height
\includegraphics[width=75mm]{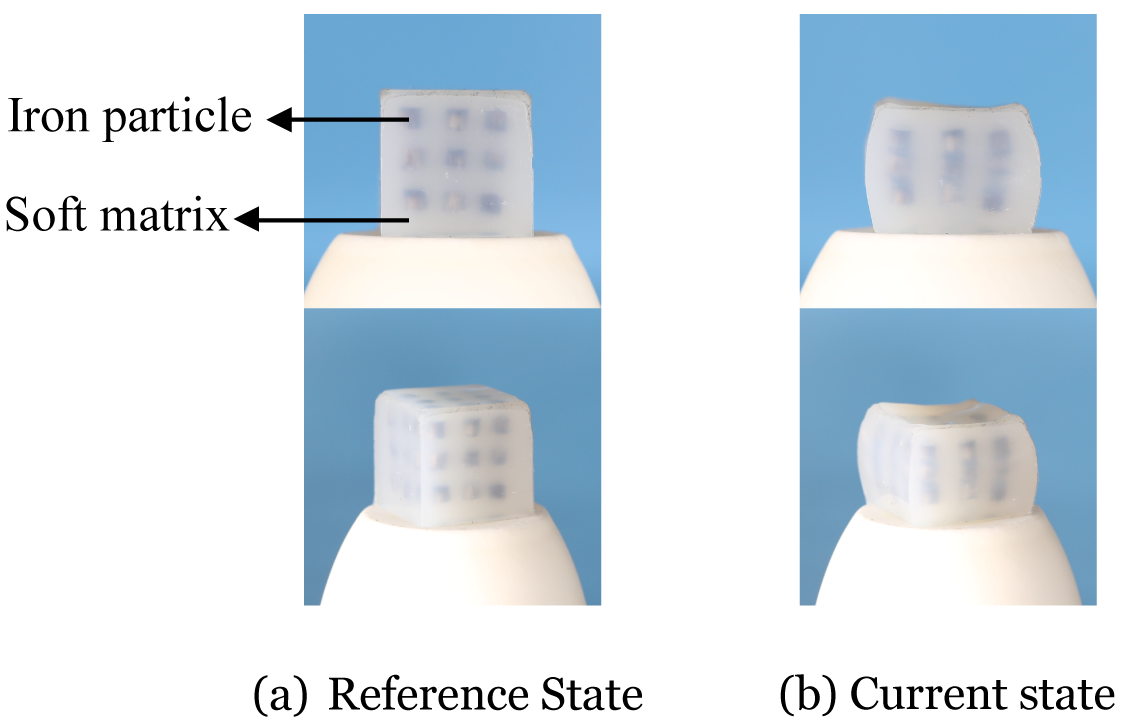}
\captionsetup{labelformat=empty} % makes sure dummy caption is blank
\caption{} % add dummy caption - otherwise \label won't work and figure numbering will not count up
\label{fig1} % use \ref{fig1} to reference to this figure
\end{wrapfigure} % avoid blank space here(a) Under a uniform vertical magnetic field, particle 1 and particle 2 are subject to attractive force if they are placed vertically, subject to repulsive force if they are placed horizontally. (b) and (c) two commonly used magnetic elastomers: random distribution and chain-like distribution. \\

uniform as possible, but imperfections are inevitable. When we increased the magnetic field, the two particles with the shortest distance pulled in first. This localized pull-in distorted the whole structure. When we increased the magnetic field a little further, other pairs of particles pulled in subsequently. In the end, all particles pulled in along the vertical direction. The distance between the top particle and the bottom particle changed by about 50\% and the overall strain of the structure is about 16\%.

\section*{Magnetoelastic microbot}

\quad \quad Giant magnetoelastic strain by magnetic pull-in enables to design useful magnetic microbots. With the designed arrangement of magnetic particles in soft matrix, the microbots are capable of different functions. In this section, we show three prototypes of untethered microbots: a robot for gripping, a robot for drug delivery and a robot for jumping. 

\newpage

\marginpar{
\vspace{.7cm} % adjust vertical position relative to text with \vspace{} - note that you can enter negative numbers to move margin caption up
\color{Gray} % this gives caption a grey color to set it apart from text body
\textbf{Figure 5. Magnetoelastic actuator.} % note that \ref{fig1} refers to the corresponding wrapfigure
(a)-(b) The magnetic gripper, (c)-(f) The magnetic drug-deliverer: (c) the sketch of the microbot; (d) the robot initially floats at the surface of the water tank; (e) a gradient magnetic field drags the robot to sink down to the target place; (f) the magnetic field further increases and the iron particles in the robot pulled in to release drug. (g)-(k)The magnetic jumper (g) reference state; (h) when a small magnetic field $H_1$ is applied, the robot deforms slightly; (i)when the magnetic field reaches a critical level $H_2$, the robot suddenly jumps off the ground; (j) the jumping robot falls down; (k) the robot returns to its initial state when the magnetic field is removed.
}
\includegraphics[width=130mm]{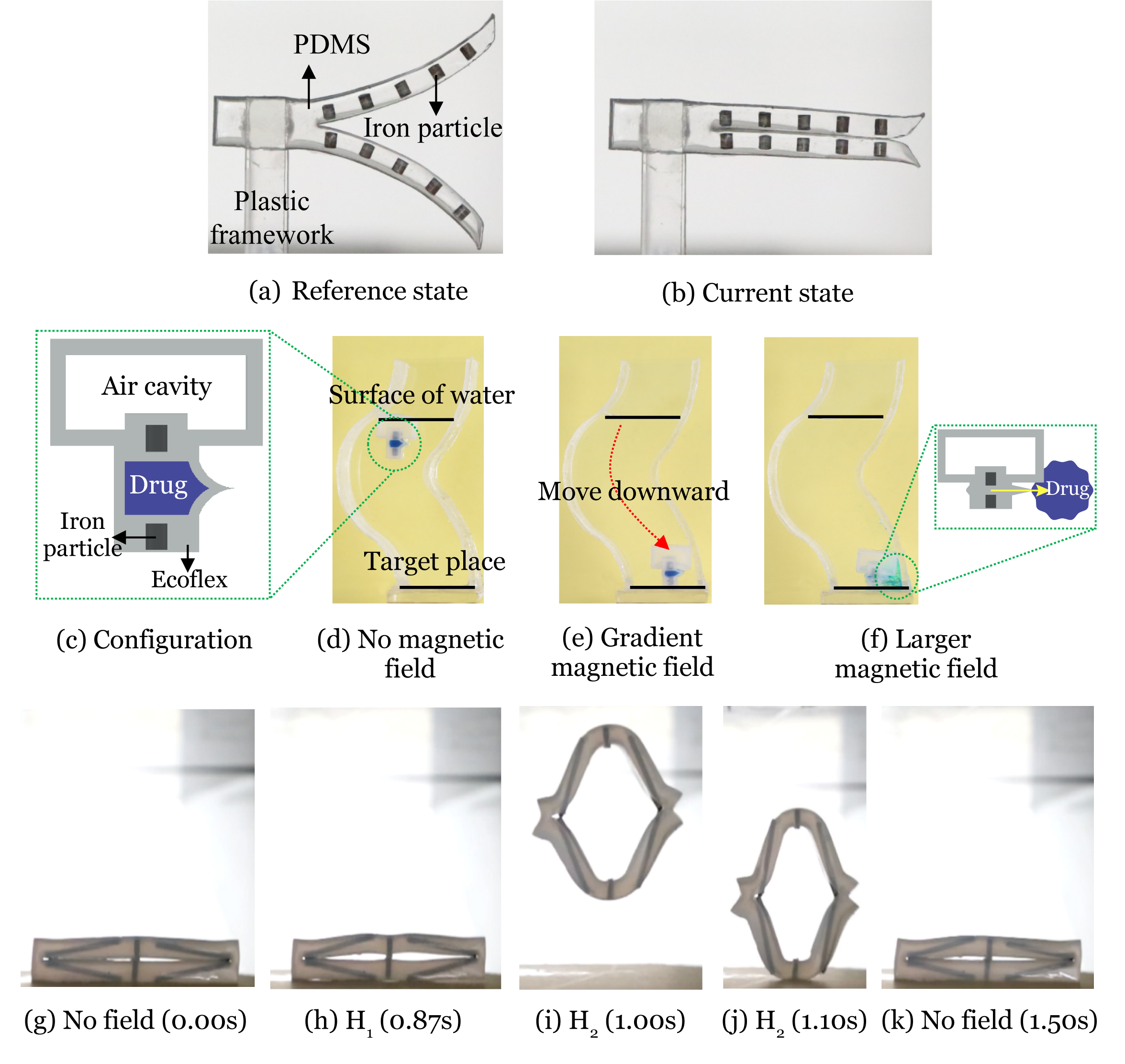}

\subsection*{Grip}

\quad \quad We poured uncured PDMS and curing agent (weight ratio=10:1) into a gripper-like mold, and placed ten iron particles inside. The PDMS matrix was cured in $60\,^{\circ}\mathrm{C}$ for 4h and then fixed to a plastic framework with super glue (Fig.5a). When a sufficiently large vertical uniform magnetic field was applied, the distance between the particles decreased and the magnetic force increased dramatically. The leftmost pair of particles with the smallest distance and the largest force pulled in first, acting like the leading wire of firecrackers (Fig.5a). It induced the other pairs to get closer in distance and pull in one by one. The pull-in of each pair of particles finally lead to a complete closure of the gripper (Fig. 5b). This design can be further programmed to adapt to the arbitrary target shape with a large gripping angle.

\subsection*{Deliver drug}

\quad \quad Fig.5c shows a microbot consisting of an elastic framework made of Ecoflex 00-30, an air cavity, a liquid chamber, and two iron particles. Iron particles were placed in the robot before the framework was cured. Blue water was injected into the liquid chamber to represent drug. The microbot floated on the surface of a water tank in the absence of magnetic field due to the buoyant force provided by the air cavity (Fig.5d). We applied a vertical uniform magnetic field superposed with a very weak magnetic field gradient. The weak field gradient overcome the buoyant force and dragged the microbot from water surface to the bottom, mimicking the guided movement in animal blood vessels (Fig.5e). When the microbot moved to the designated location, we further increased the magnitude of the uniform magnetic field so that the iron particles inside the robot pulled in and the drug was squeezed out (Fig.5f). The untethered microbot was able to perform two tasks by controlling the applied magnetic field: move along the channel and deliver drug.

\subsection*{Jump}

\quad \quad Fig.5 (g-k) shows a jumping robot driven by a uniform magnetic field. Six iron rods were placed in a soft matrix made of Ecoflex 00-30. Four long iron rods were placed diagonally and two short iron rods were placed vertically in the center. The two short rods were separated by a narrow horizontal chink (Fig.5g). When the robot was subject to a vertical magnetic field, the long rods tended to rotate to the direction towards to the external field, while the short rods tended to attract each other by pull-in effect. When the magnetic field was small, the pull-in effect of the short rods dominated compared the rotation effect of the long rods. Consequently, the robot deformed and the chink slightly opened (Fig.5h) compared to the undeformed state (Fig.5g). The short rods restrained the movement of the robot and stored the magnetic energy. When the applied magnetic field exceeded a critical value, the rotation effect of the long rods dominated, and thus the robot suddenly jumped up (Fig.5i). The energy stored in two short rods were suddenly released to accelerate the robot. The response time was roughly 0.1s. The gravity center of the robot was raised to about 8 times higher than its initial state, demonstrating a large momentum powered by magnetic pull-in.

\section*{Conclusion}

\quad \quad We report a new mechanism of magnetic pull-in instability for soft magnetic actuators. The critical condition to trigger the magnetic pull-in instability depends on the competition between the shear modulus of the matrix and the magnetic Maxwell stress. The placement of magnetic particles in the matrix greatly affects the magnetoelastic strain. We demonstrate the magnetic pull-in instability in a transparent hydrogel matrix with iron particles embedded. We further harness the magnetic pull-in to design three prototypes of untethered microbots. The first robot can grip objects with a large rotating angle. The second robot can flow in a fluid channel by magnetic navigation and perform the task of drug delivery. The third robot can self-accelerate and jump to the height 8 times of its initial height. These demonstrations show the advantages of the design soft magnetic actuators in large deformation, fast response, high power density, untetherness and robustness by the principle of magnetic pull-in instability. This mechanism may help to design versatile soft robotics towards to the applications of biomimetics and medical microbots.

\section*{Acknowledgments}
The work was supported by NSFC (No. 11772249).

\section*{Reference}

[1]. Zrinyi, M., L. Barsi and A. Büki, Ferrogel: a new magneto-controlled elastic medium. Polymer Gels and Networks, 1997. 5(5): p. 415-427.

[2]. Zrínyi, M., L. Barsi and A. Büki, Deformation of ferrogels induced by nonuniform magnetic fields. The Journal of chemical physics, 1996. 104(21): p. 8750-8756.

[3]. Zrınyi, M., et al., Direct observation of abrupt shape transition in ferrogels induced by nonuniform magnetic field. The Journal of chemical physics, 1997. 106(13): p. 5685-5692.

[4]. Zrinyi, M., D. Szabo and L. Barsi, Magnetic field sensitive polymeric actuators. Journal of intelligent material systems and structures, 1998. 9(8): p. 667-671.
 
[5]. Ramanujan, R.V. and L.L. Lao, The mechanical behavior of smart magnet–hydrogel composites. Smart materials and structures, 2006. 15(4): p. 952.
 
[6]. Snyder, R.L., V.Q. Nguyen and R.V. Ramanujan, Design parameters for magneto-elastic soft actuators. Smart Materials and Structures, 2010. 19(5): p. 055017.
 
[7]. Nguyen, V.Q., A.S. Ahmed and R.V. Ramanujan, Morphing soft magnetic composites. Advanced Materials, 2012. 24(30): p. 4041-4054.
 
[8]. Kim, J., et al., Programming magnetic anisotropy in polymeric microactuators. Nature materials, 2011. 10(10): p. 747-752.
 
[9]. Huang, H., et al., Soft micromachines with programmable motility and morphology. Nature Communications, 2016. 7.

[10]. Schmauch, M.M., et al., Chained Iron Microparticles for Directionally Controlled Actuation of Soft Robots. ACS Applied Materials \& Interfaces, 2017.

[11]. Kim, Yoonho, et al. Printing ferromagnetic domains for untethered fast-transforming soft materials. Nature, 2018, 558.7709: 274.

[12]. Li, J., et al., Micro/nanorobots for biomedicine: Delivery, surgery, sensing, and detoxification. Science Robotics, 2017. 2(4): p. eaam6431.

[13]. Nelson, B.J., I.K. Kaliakatsos and J.J. Abbott, Microrobots for minimally invasive medicine. Annual review of biomedical engineering, 2010. 12: p. 55-85.

[14]. Martel, S., Beyond imaging: Macro-and microscale medical robots actuated by clinical MRI scanners. Science Robotics, 2017. 2(3): p. eaam8119.

[15]. Rahmer, J., C. Stehning and B. Gleich, Spatially selective remote magnetic actuation of identical helical micromachines. Science Robotics, 2017. 2(3): p. eaal2845.

[16]. Chin, S.Y., et al., Additive manufacturing of hydrogel-based materials for next-generation implantable medical devices. Science Robotics, 2017. 2(2): p. eaah6451.

[17]. Dogangil, G., et al. Toward targeted retinal drug delivery with wireless magnetic microrobots. in Intelligent Robots and Systems, 2008. IROS 2008. IEEE/RSJ International Conference on. 2008: IEEE.

[18]. Zhao, X., et al., Active scaffolds for on-demand drug and cell delivery. Proceedings of the National Academy of Sciences, 2011. 108(1): p. 67-72.

[19]. Jeong, S., et al., Enhanced locomotive and drilling microrobot using precessional and gradient magnetic field. Sensors and Actuators A: Physical, 2011. 171(2): p. 429-435.

[20]. Nowak, H., Magnetism in medicine: a handbook. 2007: John Wiley \& Sons.

[21]. Fluckiger, M. and B.J. Nelson. Ultrasound emitter localization in heterogeneous media. in Engineering in Medicine and Biology Society, 2007. EMBS 2007. 29th Annual International Conference of the IEEE. 2007: IEEE.

[22]. Jeong, S., et al., Penetration of an artificial arterial thromboembolism in a live animal using an intravascular therapeutic microrobot system. Medical engineering \& physics, 2016. 38(4): p. 403-410.

[23]. Ullrich, F., et al., Mobility experiments with microrobots for minimally invasive intraocular SurgeryMicrorobot experiments for intraocular surgery. Investigative ophthalmology \& visual science, 2013. 54(4): p. 2853-2863.

[24]. Felfoul, O., et al., Magneto-aerotactic bacteria deliver drug-containing nanoliposomes to tumour hypoxic regions. Nature Nanotechnology, 2016.

[25]. Aharoni, A., Demagnetizing factors for rectangular ferromagnetic prisms. Journal of applied physics, 1998. 83(6): p. 3432-3434.

[26]. Landau, L.D., The classical theory of fields. Vol. 2. 2013: Elsevier.

[27]. Jackson, J.D., Classical electrodynamics. 2007: John Wiley \& Sons.

[28]. Filipcsei, G. and M. Zrínyi, Magnetodeformation effects and the swelling of ferrogels in a uniform magnetic field. Journal of Physics: Condensed Matter, 2010. 22(27): p. 276001.

[29]. Liao, G., X. Gong and S. Xuan, Magnetic field-induced compressive property of magnetorheological elastomer under high strain rate. Ind. Eng. Chem. Res, 2013. 52(25): p. 8445-8453.

[30]. Jolly, M.R., J.D. Carlson and B.C. Munoz, A model of the behaviour of magnetorheological materials. Smart Materials and Structures, 1996. 5(5): p. 607.

\nolinenumbers

%This is where your bibliography is generated. Make sure that your .bib file is actually called library.bib
\bibliography{library}

%This defines the bibliographies style. Search online for a list of available styles.
\bibliographystyle{abbrv}

\end{document}